\documentclass[12pt]{article}
\usepackage{amsmath}
\usepackage{amssymb}
\begin{document}
\vskip 1truein
\begin{center}
{\bf CONSISTENCE OF A GL(3,R) GAUGE FORMULATION \\
FOR TOPOLOGICAL MASSIVE GRAVITY} {\footnote {Talk given at the
Spanish Relativity Meeting 2007, Relativistic Astrophysics and
Cosmology, Puerto de la Cruz, Tenerife, Spain }}\vskip 5pt
{\bf Rolando Gaitan D.} \\
{\it Departamento de F\'\i sica, Facultad de Ciencias y
Tecnolog\'\i a,\\ Universidad de Carabobo, A.P. 129 Valencia 2001,
Edo. Carabobo, Venezuela.\\
e-mail: rgaitan@uc.edu.ve}
\end{center}
\vskip .5truein
\begin{abstract}

We include a Chern-Simons term in a $GL(3,R)$ gauge formulation of
gravity with a cosmological contribution in 2+1 dimension and we
explore consistence showing that excitations must be causal and
standard topological massive gravity is recovered from this type
of construction at the torsionless limit.

\end{abstract}



\section{Introduction}

It is well known that the introduction of a Chern-Simons
lagrangian term (CS) in the Hilbert-Einstein formulation provides
a theory which describes a massive excitation of a graviton in 2+1
dimensions[1]. If a cosmological term is included, the
cosmologically extended topological massive gravity (TMG$\lambda$)
arises[2]. The aforementioned action is
\begin{eqnarray}
S=\frac{1}{\kappa^2} \int d^3 x \sqrt{-g}(R+\lambda)+
\frac{1}{\kappa^2\mu}\,S_{CS}\,\, \,, \label{eqt1}
\end{eqnarray}
where $\kappa^2$  is in lenght units (i.e., $\kappa^2 \sim l$),
$\mu \sim l^{-1}$ and $S_{CS}$ is the CS action. In a Riemannian
space-time, the action (\ref{eqt1}) gives the field equation,
$R^{\mu \nu }-\frac{g^{\mu \nu}}2R-\lambda g^{\mu
\nu}+\frac{1}{\mu}\,C^{\mu \nu }=0$ where $C^{\mu \nu}$ is the
(traceless) Cotton tensor. The trace of the field equation gives a
consistency condition on the trace of the Ricci tensor (this
means, $R=-6\lambda$). Starting with the field equation, it is
possible to write down an hyperbolic-causal equation which
describes a massive propagation for the Ricci tensor as follows
\begin{eqnarray}
(\nabla_\mu \nabla^\mu -
\mu^2)\,R_{\mu\nu}-R^{\alpha\beta}R_{\alpha\beta}g_{\mu\nu}
+3{R^\alpha }_\mu R_{\alpha\nu} +\frac{\mu^2}{3}\,Rg_{\mu\nu}
\nonumber \\ -\frac{3}{2}\,RR_{\mu\nu} +\frac{1}{2}\, R^2
g_{\mu\nu} =0
 \, \, . \label{eqt2bb}
\end{eqnarray}

The next section is devoted to explore consistence of a $GL(3,R)$
gauge formulation[3,4]  for topological massive gravity with
cosmological constant (GTMG$\lambda$), verifying the existence of
causal propagation and the fact that standard TMG$\lambda$ can be
recovered from GTMG$\lambda$ at the torsionless limit. Some
remarks will be given in the conclusions.

\section{A  GL(3,R) gauge formulation for topological massive gravity with cosmological constant}

A brief review of the gauge formulation for (free) gravity with
cosmological constant starts here[4].  Let $M$ be a 2+1
dimensional manifold with a metric, $g_{\mu \nu }$ provided. A
(principal) fiber bundle is constructed with $M$, a 1-form
connection is given,  ${(A_\lambda)^\mu}_\nu $ which will be
though non metric dependent. The connection transforms as
${A_\lambda}^\prime =UA_\lambda U^{-1} + U
\partial _\lambda U^{-1}$ under $U \in GL(3,R)$. Torsion and curvature tensors are
${T^\mu}_{\lambda\nu}={(A_\lambda)^\mu}_\nu-{(A_\nu)^\mu}_\lambda$
and $F_{\mu\nu} \equiv D_\mu A_\nu - D_\nu A_\mu + [A_\mu , A_\nu
]$ (components of the Riemann tensor are ${R^\sigma}_{\alpha \mu
\nu }\equiv ({F_{\nu \mu})^\sigma }_\alpha$). The gauge invariant
action is
\begin{equation}
S_o = \kappa^2 \int d^3 x \sqrt{-g} \,\, (-\frac 14 \, tr\,
F^{\alpha \beta}F_{\alpha \beta} + \lambda^2 ) \,\, . \label{eqm3}
\end{equation}
which reproduces the Hilbert-Einstein with cosmological constant
field equations.

The lagrangian massive term to consider is the CS action
\begin{equation}
S_{CS}=\frac{m\kappa^2}{2}\int
d^3x\,\epsilon^{\mu\nu\lambda}\,tr\big( A_\mu \partial_\nu
A_\lambda +\frac{2}{3}\, A_\mu A_\nu A_\lambda\big)\, \, ,
\label{cs1}
\end{equation}
which is gauge variant because
\begin{equation}
\delta_U S_{CS}=-\frac{m\kappa^2}{2}\int
d^3x\,\epsilon^{\mu\nu\lambda}\,tr\,\partial_\nu \big[ A_\mu
\partial_\lambda U U^{-1}\big]-4\pi^2\kappa^2m\,W(U)
 \, \, , \label{ssm15}
\end{equation}
where $W(U)\equiv \frac{1}{24\pi^2}\int
d^3x\,\epsilon^{\mu\nu\lambda}\,tr\big( U^{-1}\partial_{\mu}
UU^{-1}\partial_{\nu} UU^{-1}\partial_{\lambda} U\big)$ is the
''winding number'' of the gauge transformation  $U$. So, the
topological massive action is
\begin{equation}
S = S_o+S_{CS} \,\,. \label{eqm10}
\end{equation}

The torsionless limit of (\ref{eqm10}) can be explored by
introducing nine constraints through the new action $S' =
S+\kappa^2\int d^3 x \sqrt{-g}
\,b_{\alpha\beta}\,\varepsilon^{\beta\lambda\sigma}{(A_\lambda)^\alpha}_\sigma$,
where $b_{\alpha\beta}$ are lagrange multipliers. Variation on
connection and metric gives rise to the following field equations
\begin{equation}
D_\mu R_{\sigma\lambda}-D_\lambda R_{\sigma\mu}
-m\,{\varepsilon^{\nu\rho}}_\sigma(g_{\lambda\nu}R_{\mu\rho}-g_{\mu\nu}R_{\lambda\rho}
-\frac{2}{3}\,Rg_{\lambda\nu}g_{\mu\rho})=0\,\, , \label{eq13}
\end{equation}
\begin{equation}
R_{\sigma\mu}{R^\sigma}_\nu
-RR_{\mu\nu}+\frac{g_{\mu\nu}}{4}\,R^2-g_{\mu\nu}\lambda^2=0\,\, ,
\label{eqa14}
\end{equation}
where the following consistency condition appears
\begin{equation}
R=constant \,\, . \label{eqam10}
\end{equation}

Due to the last condition on the Ricci scalar, we can test
solutions of the type $R_{\mu\nu}=\frac{R}{3}\,g_{\mu\nu}$, by
pluging them in (\ref{eqa14}), and this gives
\begin{eqnarray}
R=\pm 6\mid\lambda\mid\,\, , \label{eqam13}
\end{eqnarray}
verifying the existence of (Anti) de Sitter solutions.

A quick look on causal propagation of the theory can be performed
writing a second order equation from (\ref{eq13}), this means
\begin{eqnarray}
(\nabla_\alpha \nabla^\alpha -
m^2)\,R_{\mu\nu}-R^{\alpha\beta}R_{\alpha\beta}g_{\mu\nu}
+3{R^\alpha }_\mu R_{\alpha\nu} +\frac{m^2R}{3}\,g_{\mu\nu}
\nonumber \\-\frac{3R}{2}\,R_{\mu\nu} +\frac{R^2}{2}\,  g_{\mu\nu}
=0
 \, \, . \label{eq4}
\end{eqnarray}

\section{Conclusion}
Equation (\ref{eq4}) describes a massive hyperbolic-causal
propagation of graviton. So, GTMG$\lambda$ contains as a
particular case the TMG$\lambda$ classical formulation (at the
torsionless limit) if we take the mass value $m$ as the CS
($m=\mu$) and the consistency condition (\ref{eqam10}) is fixed as
(\ref{eqam13}).

Obviously, GTMG$\lambda$ is gauge variant under $GL(3,R)$ due to
the presence  of the CS term. However, by taking boundary
conditions on the elements $U$, the term $\int
d^3x\,\epsilon^{\mu\nu\lambda}\,tr\,\partial_\nu \big[ A_\mu
\partial_\lambda U U^{-1}\big]$ in (\ref{ssm15}), goes to zero
and the transformation rule now is $\delta_U
S=-4\pi^2\kappa^2m\,W(U)$. If we demand that the expectation value
of a gauge invariant operator (i.e., $<\mathcal{O}>\equiv
Z^{-1}\int \mathcal{D}A \,\mathcal{O}(A) \,e^{iS}$ with the gauge
invariant measure $\mathcal{D}A$ and the normalization constant
$Z$) must be gauge invariant too, it is required that
$-4\pi^2\kappa^2m\,W(U)$ be an integral multiple of $2\pi$ and a
quantization condition on the parameter $\kappa^2m$ must arises.
This fact occurs, at least, by performing a restriction on the
covariance of the theory, this means, taking a compact subgroup of
$GL(3,R)$ (i.e., $SO(3)$).

A first step to explore a canonical quantization program for this
theory and a study of propagation of spin degree of freedom,
between other things, could be the linearization of the metric and
connection. There, the perturbative analysis must depends non
trivially on which kind of non perturbed space-time we start
(i.e., torsionless or not). This will be studied elsewhere.

{\bf Acknowlegdment}

Author wish to thanks M. Valera for technical support. This work
is supported by Grants FONACIT G-2001000712 and CDCH-UC 1102-06.

\end{document}